\documentclass[a4paper,11pt]{article}
\usepackage{named}
\usepackage[utf8]{inputenc}
\usepackage[english]{babel}
\usepackage[a4paper]{geometry}
\usepackage{stmaryrd}
\usepackage{amsmath,bm}
\usepackage{booktabs}
\usepackage{multirow}

\newtheorem{theorem}{Theorem}

\newcommand{\lst}[2]{${#1}_0$,~${#1}_1$, $\dots\,$,~${#1}_{#2-1}$}
\newcommand{\ep}[1]{\left\llbracket #1\right\rrbracket}
\renewcommand{\epsilon}{\varepsilon}

\title{Spectral Ranking\thanks{This update integrates the \textit{Network Science} version~\protect\cite{VigSR} with Landau's 1895 work on chess tournaments.}}
\author{Sebastiano Vigna\\\large Dipartimento di Informatica\\ Università degli Studi di Milano, Italy}
\begin{document}
\bibliographystyle{named}

\maketitle

\begin{abstract}
We sketch the history of
\emph{spectral ranking}---a general umbrella name for techniques that apply the theory of
linear maps (in particular, eigenvalues and eigenvectors) to matrices that do
not represent geometric transformations, but rather some kind of
\emph{relationship between entities}. Albeit recently made famous by the ample
press coverage of Google's PageRank algorithm, spectral ranking was devised
more than a century ago, and has been
studied in tournament ranking, psychology, social sciences, bibliometrics,
economy and choice theory.
We describe the contribution given by previous scholars in precise and modern mathematical
terms: along the way, we show how to express in a general way \emph{damped}
rankings, such as Katz's index, as dominant eigenvectors of
perturbed matrices, and then use results on the Drazin 
inverse to go back to the dominant
eigenvectors by a limit process. The result suggests a 
regularized definition of spectral ranking that yields for a general matrix a
unique vector depending on a boundary condition.
\end{abstract}

\section{Introduction}

From a mathematical viewpoint, a matrix $M$ represents a linear transformation
between two linear spaces. It is just one of the possible representations of
the map---it depends on a choice for the bases of the source and target space.
Nonetheless, matrices arise all the time in many fields outside mathematics,
often because they can be used to represent (weighted) binary relations. At
that point, one can apply the full machinery of linear algebra and see what
happens. The most famous example of this kind is probably \emph{spectral graph
theory}, which provides bounds for several graph features using eigenvalues of
adjacency matrices.

When a square matrix represents relationships between entities, such as
endorsement among persons, teams defeating other teams, friends or followers on
social networks, and so on, several different eigenvectors can be obtained from
the original matrix, giving rise to different kinds of \emph{spectral rankings}.
Spectral rankings have been computed under different names since the end of the
XIX century in tournament ranking, psychology, social sciences, bibliometrics,
economy and choice theory.

This paper describes the early history of spectral ranking, highlighting
fundamental contributions. Moreover, by describing those contributions in modern
terms, we can actually describe new connections between them and show that all
rankings depending on a \emph{damping} or \emph{attenuation} factor, such as
Katz's index, are just dominant eigenvectors of perturbed matrices.
We then go back to the eigenvectors of the original matrix by a limiting
process, obtaining a regularized definition of spectral ranking that yields for
a general matrix a unique vector depending on a boundary condition.

\section{Spectral Ranking 101}

Let us start with a square matrix $M$ on the reals. We will not make any
assumption on $M$. We imagine that the indices of rows and columns actually
correspond to some entity, and that each value $m_{ij}$ represents some form
\emph{relationship} between entity $i$ and entity $j$. This relationship can 
measure a \emph{endorsement}, as in ``$i$ likes $j$ this much'', or a score of
$i$ versus $j$, as in ``$i$ beats $j$ by this amount''. The first case is common
in psychometry and sociometry, when matrices represent graphs, whereas the
second case is common in games, when matrices represent scores of
matches.

Many \emph{centrality indices} based on simple summations performed on the row
or columns of this matrix were common in psychometry and sociometry. For
instance, if the matrix contains just zeroes and ones meaning ``don't like'' or
``like'', respectively, the sum of column $j$ will tell us how many entities
like $j$. But, clearly, we are not making much progress.

In the middle of the XIX century, however, the results of chess tournaments were represented
by a square matrix $M$ populated with $0$, $1$, and $1/2$, representing a
defeat, a victory or a draw. The diagonal of such matrices was forced to be zero, and the sum of
symmetric entries would always sum to one. Row sums,\footnote{We
use row vectors.}  that is, $M\mathbf 1^T$,
were an easy way to provide a global score for a player, from which a ranking
could be derived,\footnote{Here we
take care of distinguishing the \emph{scores} given to the players from the
\emph{ranking} obtained sorting the teams by score.} so to assign prizes, or divide money proportionally. 

To improve the precision of this score, the Austrian chess player
Oscar Gelbfuhs proposed in 1873 to iterate the procedure, that is, to
compute $M^2\mathbf 1^T$, refining
previous scores. Essentially, starting from an initial
score of $1$ given to all players, each player would get a new score
obtained by adding the scores of the players that he or she defeated, and half the
scores of the players with whom there was a draw. The procedure would then be
iterated again using the new set of scores.

Edmund Landau~\cite{LanZRWT}\footnote{I must thank Jan Peter Sch\"afermeyer
for pointing me to~\cite{LanUPS}, which points to~\cite{LanZRWT}.
Previous versions of this paper~\cite{VigSR} mentioned Seeley's work as first
appearance of the idea.
To the best of my knowledge, it is still the first appearance of the
$\ell_1$-normalization idea.} noted 
in his first published paper~\cite{SchPEL} that 
if one consider the natural iterated score $M^k\mathbf 1^T$,
the results are
unreliable, as the ranking depends on the number of iterations $k$.
He thus proposed
to compute a score vector $\bm r$ satisfying, in matrix notation,
\[ M\bm r^T=\lambda\bm r^T, \]
that is, a right eigenvector (such a vector is fixed by the
refinement procedure, modulo multiplication by a constant).
The first known appearance of the idea is thus
more than a century old---in fact, it is XIX-century mathematics.

In 1895, the very existence of
such a ranking, or its properties, were unknown, but just a few years later
Oskar Perron and Georg Frobenius would develop their well-known theory of
positive matrices~\cite{BePNMMS}.
Thus, Landau returns on the subject~\cite{LanUPS} to show that
his approach gives a sensible result.\footnote{Incidentally, he includes an
example coming from correspondence with Perron showing that small fluctuations
in the matrix entries can generate paradoxical results, thus, in fact,
dismissing the method~\cite{DaEARHS,CMSFLRPV}.}

The other fundamental step towards spectral ranking was made half a century
later by John R.~Seeley~\cite{SeeNRI}, who was unaware of Landau's work: he
noted that indices based on row or column sums were not really meaningful because they were not taking into consideration
that it is important being liked by someone that is in turn being liked a lot, and so on. In other words, an index of importance, centrality, or authoritativeness, should be defined
\emph{recursively}\footnote{The word ``recursively'', here, is used in a generalized but slightly
incorrect sense which is common in the literature,
and for which the word ``impredicatively'' would be more appropriate, but probably obscure to most readers.}
 so that my score is equal to the weighted sum of the scores
of the entities that endorse me. In matrix notation,
\begin{equation}
\label{eq:seeley}
\bm r = \bm r M.
\end{equation}
Of course, this is not always possible. Seeley, however, considers a nonnegative
matrix without null rows and normalizes its rows so that they have unit $\ell_1$
norm (e.g., you divide each entry by its row sum); his rows have always
nonzero entries, so this is always possible, and Equation~\ref{eq:seeley} has a
solution, because $M\mathbf 1^T=\mathbf 1^T$, so $1$ is an eigenvalue of $M$,
and its left eigenvector(s) provide solutions to Equation~\ref{eq:seeley}.
Uniqueness is a more complicated issue which Seeley does not discuss and which
can be easily analyzed using Perron--Frobenius theory, which also shows that $1$ is the spectral
radius, so $\bm r$ is a dominant\footnote{A \emph{dominant eigenvalue} is
an eigenvalue with largest modulus (i.e., the spectral radius). An eigenvector
associated with the dominant eigenvalue is called a \emph{dominant eigenvector}.
In most practical cases of spectral ranking there is just one strictly dominant
eigenvalue.} eigenvector, and that there are
positive solutions.\footnote{Actually, Seeley exposes the entire matter in terms of linear equations. Matrix calculus is used only for solving a linear system by Cramer's rule.}

It is interesting to note that the motivation of the two works are very
different: Landau is interested in the limit of an iterative process refining a
score, and defines such a limit recursively; Seeley want to define directly a
recursive score. 

Teh-Hsing Wei in his Ph.D.~dissertation~\cite{WeiAFRT} will duplicate
Landau's work, unaware of his results. Kendall~\cite{KenFCTPC} will discuss
Wei's (unpublished) findings at length, and as a result the eigenvector-based 
techniques proposed by Landau will become known in the literature about ranking
of sport teams as ``Perron ranking'' or ``Kendall--Wei ranking''.

Few years later, in his celebrated book on graph
theory Claude Berge quotes Wei's thesis and makes the remark that the (right)
eigenvector approach can be applied to \emph{any} directed graph. As an example, he considers \emph{sociograms}, 
where nodes represent individuals and arcs represent
influence~\cite{BerTGA}.\footnote{Indeed, if we had to assign a name to
the dominant eigenvector it should be ``Landau--Berge index'' or 
``Landau--Berge centrality''.}

Getting back to left eigenvectors, the works of Landau, Seeley and Berge
suggest that we consider matrices $M$ with a real and positive dominant 
eigenvalue $\lambda$ and its eigenvectors, that is, vectors $\bm
r$ such that
\begin{equation}
\label{eq:seeley2}
\lambda\bm r = \bm r M.
\end{equation}
If $\lambda$ is complex, $\bm r$ cannot be real, and the lack of an ordering
that is compatible with the field structure makes complex numbers a bad
candidate for ranking.

In general, a \emph{(left)\footnote{The distinction between left and right
spectral ranking is in principle, of course, useless, as the left spectral
ranking of $M$ is the right spectral ranking of $M^T$.
Nonetheless, the kind of motivations leading to the two kind of rankings are quite different, and we feel that it is useful to keep around the distinction: if the matrix represents \emph{endorsement}, left spectral ranking is the correct choice; if the matrix
represents \emph{influence} or ``better-than'' relationships right spectral ranking should be used
instead.} spectral ranking} associated with $M$ is a dominant (left) eigenvector.
If the eigenspace has dimension one, we can speak of \emph{the} spectral ranking
associated with $M$. Note that in principle such a ranking is defined up to a
constant: this is not a problem if all coordinates of $\bm r$ have the same sign,
but introduces an ambiguity otherwise.

\section{Damping}

We will now start from a completely different viewpoint. If the matrix $M$ is a
zero/one matrix, the entry $i,j$ of $M^k$ contains the number of directed path
from $i$ to $j$ in the directed graph defined by $M$ in the obvious way. A
reasonable way of measuring the importance of $j$ could be measuring the number
of paths going into $j$, as they represent recursive endorsements.\footnote{Indeed, taking the limit of the 
vector giving for each node the number incoming paths of length $k$ (somehow normalized) when $k\to\infty$
leads to the definition of spectral ranking, as the process is equivalent to 
finding the dominant left eigenvector using the power method. This was observed
already by Berge~\cite{BerTGA}.} Unfortunately, trying the
obvious, that is,
\[
\bm 1 (1 + M + M^2 + M^3 + \cdots ) = \bm 1\sum_{k=0}^\infty M^k
\]
will not work, as formally the above equation is correct, but convergence is
not guaranteed. Convergence is guaranteed, however, if $M$ has spectral radius
smaller than one, that is, $|\lambda_0|<1$. It is thus tempting to introduce an
\emph{attenuation} or \emph{damping} factor that makes things work:
\begin{equation}
\label{eq:katz}
\mathbf 1 (1 + \alpha M + \alpha ^2M^2 + \alpha ^3M^3 + \cdots ) = \mathbf
1\sum_{k=0}^\infty (\alpha M)^k
\end{equation}
Now we are actually working with $\alpha M$, which has spectral radius smaller
than one as long as $\alpha<1/|\lambda_0|$ (e.g., if $M$ is (sub)stochastic any
$\alpha<1$ will do the job). This index was proposed by Leo
Katz~\cite{KatNSIDSA}.\footnote{We must note that actually Katz's index is
$\mathbf 1M\sum_{k=0}^\infty (\alpha M)^k$. This additional multiplication by $M$ is somewhat common in the literature; it is probably a case of \textit{horror vacui}.} He notes that
\[
\mathbf 1\sum_{k=0}^\infty (\alpha M)^k= \mathbf 1(1-\alpha M)^{-1},
\]
which means that his index can be computed solving the linear
system
\[
\bm x(1-\alpha M) = \mathbf 1.
\] 

\section{Boundary conditions}

There is still an important ingredient we are missing: some \emph{initial
preference}, or \emph{boundary condition}, as Hubbell~\cite{HubIOACI} calls
it.
Hubbell's interest is \emph{clique detection}, an early study of \emph{spectral
graph clustering}.\footnote{It would be interesting to write a note similar to
this one for spectral clustering, as sociologists have been playing with the
idea for quite a while.} Hubbell is inspired by the works of Luce, Perry and
Festinger on clique identification~\cite{LuPMMAGS,FesASUMA}; they use fixed
powers of the adjacency matrix to estimate the similarity of nodes, and Hubbell
proposes to sum up \emph{all powers} of a matrix when such a sum exists. Then,
in analogy with Leontief's input-output economic model,\footnote{Recently,
Franceschet~\cite{FraPR} has argued that Leontief's input-output model is a
precursor of PageRank.
I think this is a red herring, as Leontief just wants to represent the
relationship between input and output of an economy. He claims that an
\emph{equilibrium} is reached when prices are given by the fixpoint of the
linear operator describing the input/output relationship, but being the goods
indexing the matrix inhomogeneous, this pricing is not a ranking (and, indeed,
Leontief does not appear to make claims in this direction).} which represents
the relationships between input and output of goods in each
industry~\cite{LeoSAE}, he argues that one can define a status index $\bm r$
using the recursive equation
\begin{equation}
\label{eq:hubbell}
\bm r = \bm v + \bm r M,
\end{equation}
where $\bm v$ is a \emph{boundary condition}, or \emph{exogenous contribution}
to the system. Finally, he notes that formally  
\[
\bm r = \bm v  ( 1 - M )^{-1} = \bm v \sum_{k=0}^\infty M^k,
\]
and that the right side converges as long as $|\lambda_0|<1$: $M$ can even have
negative entries. Clearly
this is a generalization of Katz's index\footnote{Hubbell claims that its
index (actually, its \emph{status model}) bears a ``rough resemblance'' to
Katz's: once the mathematics has been laid out in simple terms, one can easily
see that they are the same thing.} to general matrices that adds an initial
condition, as the vector $\bm 1$ is replaced by the more general boundary condition 
$\bm v$.\footnote{We note that while the ranking induced 
by $\mathbf 1 M(1-\alpha M)^{-1}$ and $\mathbf 1(1-\alpha M)^{-1}= \mathbf 1 +
\mathbf 1 \alpha M(1-\alpha M)^{-1}$ is the same, this is no longer true when we
use a general boundary condition.}\footnote{Hubbell remarks that its score depends linearly on the 
boundary condition. This is actually an important feature for quick
computation of personalized or topical versions of PageRank~\cite{JeWSPWS}.}

\section{From eigenvectors to path summation}

Landau's, Seeley's and Katz's work might seem unrelated.
Nothing could be farther from truth. Let's get back to the basic
spectral ranking equation:
\[
 \lambda_0 \bm r = \bm r M.
\]
When the eigenspace of $\lambda_0$ has dimension larger than one, there is no
clear choice for $\bm r$. But we can try to \emph{perturb} $M$ so that this
happens. A simple way is using Brauer's results~\cite{BraLCRMIV} about
eigenvector separation:\footnote{I learned
the usefulness of Brauer's results in this context for separating eigenvalues
from Stefano Serra--Capizzano. The series of papers by Brauer is also (maybe
not surprisingly) quoted by Katz in his paper~\cite{KatNSIDSA}.}
\begin{theorem}
Let $A$ be an $n\times n$ complex matrix, \lst\lambda n be the eigenvalues of
$A$, and let $\bm x$ be a nonzero 
complex vector such that $A\bm x^T = \lambda_0 \bm x^T$. Then, for every complex vector $\bm
v$, the eigenvalues of $A + \bm x^T \bm v$ are 
$\lambda_0 + \bm v \bm x^T$,~$\lambda_1$, $\dots\,$,~$\lambda_{n-1}$.
\end{theorem}

Brauer's theorem suggests to perform a rank-one convex perturbation of $M$
using a vector $\bm v$ satisfying $\bm v\bm x^T=\lambda_0$ by applying the
theorem to $\alpha M$ and $(1-\alpha)\bm x^T\bm v$:
\[
 \lambda_0 \bm r = \bm r (\alpha M+ (1-\alpha)\bm x^T\bm v).
\]
Now $\alpha M+ (1-\alpha) \bm x^T\bm v$ has the same dominant eigenvalue
of $M$, but with algebraic multiplicity one, and all other eigenvalues are
multiplied by $\alpha$. This ensures that we have a unique $\bm r$, at the
price of having introduced a parameter (the choice of $\bm x$ is particularly
simple in case $M$ is stochastic, as in that case we can take $\mathbf 1$).

There is also another important consequence: $\bm r$ is defined up
to a constant, so we can impose that $\bm r\bm x^T=\lambda_0$ (i.e., in case
$\bm x=\mathbf1$, that the sum of $\bm r$'s coordinates is $\lambda_0$, which
implies, if all coordinates have the same sign, that $\|\bm r\|_1=\lambda_0$).
We obtain
\[
 \lambda_0 \bm r= \alpha\bm r M + (1-\alpha)\lambda_0\bm v,
\]
so now 
\begin{equation}
\label{eq:perturb}
\bm r = (1-\alpha)\bm v \bigl( 1- \alpha M/\lambda_0\bigr)^ {-1} = (1-\alpha) \bm
v \sum_{k=0}^\infty\bigl(\alpha M/\lambda_0\bigr)^k =(1-\lambda_0\beta)
\bm v \sum_{k=0}^\infty(\beta M)^k ,
\end{equation}
and the summation certainly converges if $\alpha< 1$ (or,
equivalently, if $\beta < 1/\lambda_0$). In other words, Katz--Hubbell's index
(modulo a multiplicative constant) can be obtained as the spectral ranking of a
rank-one perturbation of the original matrix.\footnote{This result is of course
well known for stochastic matrices, as it is one of the ways of defining
PageRank.} As the convex perturbation parameter $\alpha$ moves from $0$ to
$1$, the damping factor $\beta$ moves from $0$ to $1/\lambda_0$.

\section{From path summation to eigenvectors}

A subtler reason takes us backwards. Given a matrix $Z$, the \emph{index} of $Z$
is the smallest nonnegative integer $k$ such that the kernel (the eigenspace associated to the eigenvalue zero) of
$Z^k$ is equal to the kernel of $Z^{k+1}$. Equivalently, the index is the
size of the largest Jordan block of the eigenvalue zero (which is zero for a nonsingular matrix).

The \emph{Drazin inverse} of $Z$, written $Z^D$, is the unique solution of the equations
\[
ZXZ=X\qquad XZ=ZX\qquad Z^{\nu+1}X=Z^\nu,
\]
where $\nu$ is the index of $Z$~\cite{DraPIARS}. When $Z$ is nonsingular, it coincides with
$Z^{-1}$. The matrix $\ep Z=1-ZZ^D$ is called the
\emph{eigenprojection} (for the eigenvalue zero) of $Z$: it is a projection on the kernel of $Z^\nu$ along the range
of $Z^\nu$. The fundamental theorem
we will use about the Drazin inverse is due to Meyer~\cite{MeyLISQ}:
\begin{theorem}
Let $Z$ be a square matrix with index $\nu$. If $m$ and $p$ are nonnegative integers, the limit
\[
\lim_{\epsilon\to0}\epsilon^m(\epsilon + Z)^{-1}Z^p
\] 
exists if and only if $m+p\geq\nu$, and in that case
\[
\lim_{\epsilon\to0}\epsilon^m(\epsilon + Z)^{-1}Z^p =
\begin{cases}
Z^DZ^p & \text{if $m=0$;}\\
(-1)^{m-1}Z^{m+p-1}\ep Z & \text{if $m>0$.}
\end{cases}
\]
\end{theorem}

We would like to know what happens to the perturbed dominant eigenvector~(\ref{eq:perturb}) when $\alpha\to1$:  
\begin{align*}
\lim_{\alpha\to 1}(1-\alpha)\bm v \bigl( 1- \alpha M/\lambda_0\bigr)^ {-1}
= & \lim_{\alpha\to 1}\frac{1-\alpha}\alpha\bm v \biggl( \frac1\alpha - M/\lambda_0\biggr)^ {-1}\\ 
= & \lim_{\alpha\to 1}\frac{1-\alpha}\alpha\bm v \biggl( \frac1\alpha - 1 + 1 - M/\lambda_0\biggr)^ {-1}\\ 
= & \lim_{\alpha\to 1}\frac{1-\alpha}\alpha\bm v \biggl( \frac{1-\alpha}\alpha + \bigl(1 - M/\lambda_0\bigr)\biggr)^ {-1}\\ 
= & \lim_{\epsilon\to 0}\epsilon\bm v \bigl( \epsilon + \bigl(1 - M/\lambda_0\bigr)\bigr)^ {-1}. 
\end{align*}
Now, $1-M/\lambda_0$ is a singular matrix with index equal to the
index\footnote{The index of the eigenvalue $\lambda$ of a matrix $M$ is the
index of the matrix $1-M/\lambda$, or equivalently the maximal dimension of the
Jordan blocks of $M$ containing $\lambda$.} of $\lambda_0$ in $M$.
If $\lambda_0$ is \emph{semisimple} (i.e., $\nu=1$) we can apply Meyer's theorem with $m=1$, $p=0$ and conclude that
\[
\lim_{\alpha\to 1}(1-\alpha)\bm v \bigl(1 - \alpha M/\lambda_0\bigr)^ {-1} =\bm v \ep{1-M/\lambda_0}.
\] 
The circle is closed, as the kernel of $1-M/\lambda_0$ is exactly the eigenspace of $M$ associated with the eigenvalue $\lambda_0$: 
spectral ranking is just the limit case of Katz--Hubbell's index.\footnote{Once
again, this result is well known for PageRank~\cite{BSVPFD}.}

The eigenprojection $\ep{1-M/\lambda_0}$ has an intuitive description if $\lambda_0$ \emph{and all other eigenvalues of maximum modulus} are semisimple: 
it is equal~\cite{RotESMP} to the \emph{Ces\`aro limit}
\[
\lim_{n\to\infty}\sum_{k=0}^{n-1}\frac{(M/\lambda_0)^k}n,
\]
that is, the limit in average of $(M/\lambda_0)^n$. In stochastic matrices all eigenvalues of modulo one are semisimple, which
explains why the Ces\`aro limit is very popular in the Markov chain literature to obtain eigenprojections.
Rothblum shows~\cite{RotESMP} that a suitable generalization of the
Ces\`aro limit to \emph{$n$-fold sums} makes it possible to write the eigenprojections in limit form even if the eigenvalues of maximum modulus other than $\lambda_0$ are not semisimple.

If $\lambda_0$ is not semisimple, say with index $\nu$, the situation is not very different: 
Meyer's theorem essentially tells us that the perturbed dominant eigenvector~(\ref{eq:perturb})  
would grow too quickly as $\alpha\to 1$, but nonetheless
\[
\lim_{\alpha\to 1}(1-\alpha)^\nu\bm v \bigl( 1- \alpha M/\lambda_0\bigr)^ {-1} =(-1)^{\nu-1}\bm v \ep{1- M/\lambda_0} \bigl( 1- M/\lambda_0\bigr)^{\nu-1}.
\]
The resulting vector is still in the kernel of $1-M/\lambda_0$, because $\ep{1- M/\lambda_0}$ projects on the kernel of $\bigl( 1- M/\lambda_0\bigr)^\nu$:
\[
\Bigl(\bm v \ep{1- M/\lambda_0} \bigl( 1- M/\lambda_0\bigr)^{\nu-1}\Bigr)\bigl( 1- M/\lambda_0\bigr) = 
\bm v \ep{1- M/\lambda_0} \bigl( 1- M/\lambda_0\bigr)^\nu = 0.
\]
Note that since the $(1-\alpha)^{\nu}$ factor is just a scalar, we can also write (even when $\nu=1$)
\[
\lim_{\alpha\to 1}\frac{\bm v \bigl( 1 - \alpha M/\lambda_0\bigr)^ {-1}}{\Bigl\|\bm v \bigl( 1 - \alpha M/\lambda_0\bigr)^ {-1}\Bigr\|} =
\frac{(-1)^{\nu-1}\bm v \ep{1- M/\lambda_0}\bigl( 1- M/\lambda_0\bigr)^{\nu-1}}{\Bigl\|\bm v \ep{1- M/\lambda_0}\bigl( 1- M/\lambda_0\bigr)^{\nu-1}\Bigr\|},
\]
which means that the \emph{direction} of the perturbed dominant eigenvector always tends to that of a dominant eigenvector of $M$ as $\alpha\to1$, albeit normalization is necessary to avoid divergence.
If $M$ is nonnegative, Rothblum shows that generalized Ces\`aro limits can be applied even to this case.

\section{Putting It All Together}
\label{sec:putting}

It is interesting to note that the journey made by our original definition
through perturbation and then limiting has an independent interest. We started
with a matrix $M$ with possibly many eigenvectors associated with the dominant
eigenvalue, and we ended up with a \emph{specific} eigenvector associated with
$\lambda_0$, given the boundary condition $\bm v$. This suggests to define in
general \emph{the} spectral ranking\footnote{We remark that in social sciences
and social-network analysis ``eigenvector centrality'' is often used to name
collectively ranking techniques using eigenvectors (``centrality'' is the
sociologist's ``ranking''). On the other hand, in those areas indices
based on paths such as Katz's are considered to be different beasts.}
of $M$ with boundary
condition $\bm v$ when the dominant eigenvalue $\lambda_0$ is semisimple as
\[
\bm r =\bm v\ep{1-M/\lambda_0},
\]
or, in the general case, 
\begin{equation}
\label{eq:general}
\bm r =\bm v(-1)^{\nu-1}\ep{1- M/\lambda_0} \bigl( 1- M/\lambda_0\bigr)^{\nu-1},
\end{equation}
with $\nu$ equal to the index of $\lambda_0$.
If $M$ has a unique eigenvector, this definition is equivalent
to~(\ref{eq:seeley2}), and $\bm v$ is immaterial. However, in the general case~(\ref{eq:general})
provides a unique (albeit possibly difficult to compute) eigenvector depending
on $\bm v$.

If we start from a generic nonnegative matrix $M$ and assume to normalize its rows, obtaining a
substochastic matrix $P$, we should probably speak of \emph{Markovian spectral
ranking}, as the Markovian nature of the object becomes dominant. If $\lambda_0=1$ we have
\begin{equation}
\label{eq:markovian}
\bm r =\bm v\ep{1-P}=\bm v\lim_{n\to\infty}\sum_{k=0}^{n-1}\frac{P^k}n,
\end{equation}
as dictated by Markov chain theory.\footnote{Substochastic matrices, too, enjoy the property
that all eigenvalues of modulo one, if any, are semisimple; however, if $\lambda_0<1$ there are no such eigenvalues.} If $\bm v$ is a distribution, $\bm r$ is
essentially\footnote{``Essentially'' because $\ep{1-P}$ smooths out problems
due to periodicities in the matrix.} the limit distribution when the chain is started
with distribution $\bm v$. If $\lambda_0<1$, we must check its index and work the
details as in the general case~(\ref{eq:general}), as we can no longer guarantee
semisemplicity of all eigenvalues of maximum modulus. For example, the matrix
\[
\left(\begin{matrix}
0 & \frac12 & \frac12 \\
0 & 0 & 1\\
0 & 0 & 0
\end{matrix}\right)
\]
has a single eigenvalue $0$ with algebraic multiplicity three and geometric multiplicity one.

Finally, we could define the \emph{damped spectral ranking} of $M$
with boundary condition $\bm v$ and damping factor $\alpha $ when the dominant
eigenvalue $\lambda_0$ of $M$ is semisimple as
\[
\bm r_\alpha = 
(1-\alpha)\bm v\sum_{k=0}^\infty\bigl(\alpha M/\lambda_0\bigr)^k
\]
for $|\alpha|<1$. If we normalize the rows of $M$, we should speak of
a \emph{damped Markovian spectral ranking}.

The $(1-\alpha)$ term comes out naturally
from~(\ref{eq:perturb}), and makes the limit 
for $\alpha\to1$ exist in the semisimple case, unifying the treatment of
normalized (i.e., Markovian) and unnormalized rankings; moreover, it forces $\|\bm r\|_1=1$
when $M$ is stochastic and $\bm v$ is a distribution.\footnote{As noted by
Bonacich~\cite{BonPC}, $\alpha$ can even be negative.} An even more
general definition, taking the index $\nu$ of $\lambda_0$ into consideration, would
use a multiplicative factor $(1-\alpha)^\nu$, which would make the limit
for $\alpha\to1$ always exist. It can be argued, however, that for practical
purposes the normalization factor is just a nuisance, as it does not affect
ranks (just scores), and it might be difficult to determine, as it depends on
the index of $\lambda_0$.

It is interesting to note that in the Markovian case the change of role of the
boundary condition from the damped to the non-damped case has a simple
interpretation: in the damped case, we have a Markov chain that \emph{restarts} 
on a fixed distribution $\bm v$,
and there is a single stationary distribution which
is the limit of \emph{every} starting distribution; in the non-damped
case, $\bm v$ is the starting distribution from which we compute the limit distribution using the eigenprojection. 
Thus, when $\alpha\to1$, the restart distribution $\bm v$ becomes the starting
distribution, which is significant only if the graph underlying
the chain has more than one terminal strongly connected component. 
An analogous consideration
can be done in the general semisimple case: the preference vector of Katz--Hubbell's index
becomes, in the limit, the starting vector of an averaged version of the power method
that mimics Ces\`aro limit.

\begin{table}
\caption{\label{table}A reasoned table of the basic types of spectral ranking; the names defined
in Section~\ref{sec:putting} are in boldface.}
\centering
\begin{tabular}{lcc}
\\[1em]
 			& No normalization & Row normalization\\
\hline
\multirow{3}{*}{No damping} & \textbf{Spectral ranking} & \textbf{Markovian spectral ranking}\\
& Eigenvector centrality & Markov chain steady state\\
 & \cite{LanZRWT,BerTGA} & \cite{SeeNRI}\\[1em]
\multirow{3}{*}{Damping} & \quad\textbf{Damped spectral ranking} & \quad\textbf{Damped Markovian spectral ranking} \\
 & Katz's index & Total effect centrality, PageRank \\
 & \cite{KatNSIDSA} & \cite{FriTFCM,PBMPCR} \\
\hline
\end{tabular}
\end{table}

%
%

\section{Followers}
\label{sec:followers}

The work of Landau was forgotten, Seeley's paper was almost unnoticed, Berge's
observations were buried in a book on graph theory, and Katz's paper was known
mainly by sociologists, so it is no surprise that spectral ranking has been
rediscovered several times.

In this section we gather the main insurgencies of spectral
ranking in various fields we are aware of. In some cases, spectral ranking in
some form is applied to some domain; in other cases, very mild variations of
previous ideas are proposed (mostly, we must unhappily say, without motivation
or assessment). The list is certainly incomplete.

\noindent\textbf{\cite{FreFTSP}} For completeness, we mention French's theory of
social power, which bears a superficial formal resemblance with spectral
ranking. However, French's theory considers \emph{right} eigenvectors of
row-stochastic matrices, so the trivial uniform solution is always a solution,
and it is considered a \emph{good} solution, as the theory studies the formation
of consensus (e.g., the probability of getting the trivial uniform solution
depending on the structure of the graph).

\noindent\textbf{\cite{GouGIE}} Gould proposes to use eigenvectors to
study geographical features such as irregular terrains, the orientation of
transportation networks, and even the homogeneity of architectural features. In
one particular example, he consider the connectivity graph on a set of cities
(edges are direct roads) and discusses the information conveyed by the dominant
and a few other eigenvectors (Landau's, Seeley's and Berge's
work are not quoted).

\noindent\textbf{\cite{BonFWASSCI}} Bonacich proposes to
use spectral ranking on zero-one matrices representing entities and
their relationships to identify the most important entities (Landau's, Seeley's
and Berge's work are not quoted).

\noindent\textbf{\cite{PiNCIJASP}} Here $M$ is the matrix that contains in
position $m_{ij}$ the number of references from journal $j$ to journal $i$. The
matrix is then normalized in a slightly bizarre way, that is, by dividing $m_{ij}$ by
the $j$-th [sic] row sum. The spectral ranking on this matrix is then used to
rank journals. \cite{GelCIMPN} tries to bring Markov-chain theory in by
suggesting to divide by the $i$-th row sum instead (i.e., Markovian spectral
ranking). No reference is made to previous work.

\noindent\textbf{\cite{HoeNSSASN}} Hoede proposes to avoid the boundary
condition of Hubbell's index by computing $\bm 1 M ( 1 - M )^{-1}$ instead, under the
condition that $1-M$ is invertible. This is exactly Katz's index with no
damping. The main point of the author is that now we can just tweak the entries
of $M$ so to make $1-M$ invertible, as ``this hardly influences the model''
[sic].

\noindent{\textbf{\cite{SaaAHP}}} In the '70s, Saaty developed the theory
of the \emph{analytic hierarchy process}, a structured technique for
dealing with complex decision. After some preprocessing, a table comparing
a set of alternatives pairwise is filled with ``better then'' values (the entry
$m_{ij}$ means how much $i$ is better than $j$, and the matrix must be
reciprocal, i.e., $m_{ij}=1/m_{ji}$); \emph{right} spectral ranking is then used
to rank the alternatives. Some insight as to why this is sensible can be found
in~\cite{SaaRAP}. The mathematics is of course identical to Landau's and
Berge's, as the motivation is structurally similar.

\noindent\textbf{\cite{BonPC}} Bonacich proposes a mild extension of Katz's index
(i.e., damped spectral ranking) that includes negative damping; the interpretation
proposed is that in \emph{bargaining} having a powerful neighbor should count
negatively.

\noindent\textbf{\cite{BonSGIC}} Bonacich proposes to rank individuals and groups
simultaneously, given the person/group incidence matrix, by computing the first
left and right singular vectors of the incidence matrix $M$, which are
the spectral rankings of $MM^T$ and $M^TM$, respectively.

\noindent\textbf{\cite{FriTFCM}} Friedkin discusses a damped version of French's
consensus theory based on a row-stochastic matrix $M$. However, halfway through
the paper he defines the \emph{total effect centrality} of an entity as the average of the
off-diagonal entries of a column of $(1-\alpha)(1-\alpha M)^{-1}$.
If we forget the motivations behind Friedkin's model and take the average
of \emph{all} entries, instead, we obtain a damped (Markovian) spectral ranking
of $M$ with constant boundary condition. If $M$
is obtained by normalizing the rows of the adjacency matrix of a graph, 
we obtain exactly PageRank~\cite{PBMPCR}.

\noindent\textbf{\cite{KeePFTRFT}} Keener discusses a few different rankings for
football teams based on the results of played games. He computes dominant
eigenvectors of different matrices derived from game outcomes.
This paper is one of the very few ones explicitly quoting both Wei and Berge,
albeit not Landau.

\noindent\textbf{\cite{PBMPCR}} PageRank is the damped Markovian spectral
ranking of the adjacency matrix of a web graph (i.e., the graph having pages as
nodes and hypertextual links as arcs).
It was one the features originally used by Google to rank documents\footnote{It
should be noted there are several other ways to use a graph to obtain
scores for documents. For instance, one can use links (in particular, hypertext
links) to alter text-based scores using the score of pointed pages: this simple
idea dates at least back to the end of the eighties~\cite{FriSIHMH}. In the
nineties, the idea was rediscovered again for the web: see, for example,
\cite{MarQCIW}, which, in spite of some claims floating around the net,
does not do any kind of spectral ranking. An obvious spectral approach would
use a preference vector containing normalized text-based
scores, and then a right or left spectral ranking depending on whether
authoritativeness or relevance is to be scored. To the knowledge of the
author, this approach has not been explored yet.} associated with a query, and it has
since then been applied to a wide variety of problems~\cite{GlePBW}.
The boundary condition 
is called \emph{preference vector}, and it can be used to bias PageRank with
respect to a topic or to personal preferences. In case the preference vector is constant, the resulting
centrality is almost identical to the \emph{total effect centrality}
defined by Friedkin~\cite{FriTFCM}, whose work is not
quoted, though.

\noindent\textbf{\cite{HPPSRWWWS}} With the aim of predicting the number of
visits to a web page, Huberman, Pirolli, Pitkow and Lukose study a model derived
from \emph{spreading activation networks}. Essentially, given a distribution $d$
that tells which fraction of surfers are still surfing after time $t$,
the prediction vector at time $t$ is $d(t)\bm v P^t$, where $\bm v$ is the initial number of
surfers at each page. They use an inverse Gaussian distribution obtained
experimentally, but using a geometric distribution the predicted overall
(i.e., summed up over all $t$) number of surfers at each page will give a
Markovian damped spectral ranking.

\noindent\textbf{\cite{KleASHE}} HITS is Kleinberg's algorithm for
finding \emph{authorities} and \emph{hubs} in a (part of a) web graph.
As in Bonacich's previous work~\cite{BonSGIC},
HITS computes the first left and right singular vectors of a matrix $A$, which
are the spectral ranking of $AA^T$ and $A^TA$, respectively. Kleinberg notes,
however, that it is possible to extract clustering
information from additional singular vectors.

\noindent\textbf{\cite{BoLEMCAR}} Bonacich and Lloyd propose again to use
damped spectral ranking, but with a boundary
condition, that is, Katz--Hubbell's index.
The authors do quote Hubbell~\cite{HubIOACI}, but apparently
they do not realize that they are just redefining his index. They prove,
however, prove that under strong conditions ($M$ symmetric and with a strictly dominant eigenvalue) damped spectral
ranking converges to spectral ranking.


\noindent{\textbf{\cite{BWWEM}}} \emph{Eigenfactor} is a score computed
to score journals. It is a Markovian damped spectral ranking
computed on the citation matrix, with an additional 
non-damped step (e.g., $S(1-\alpha
S)^{-1}$), as in the original formulation of Katz's index.


\section{Acknowledgments}

The author would like to thank David Gleich for many useful discussions and
comments, and Jan Peter Sch\"afermeyer for pointing out~\cite{LanUPS}.

\section{Conclusions}

We described a comprehensive framework for spectral
ranking, highlighting the fundamental contributions of Landau, Seeley
and Berge (the dominant eigenvector, possibly with stochastic normalisation), Katz (damping) and Hubbell
(boundary condition). We showed in a very general setting that one can move
from dominant eigenvectors to damped rankings by a perturbation process, and
get back to the dominant eigenvectors by taking the limit of the parameter
controlling the perturbation. The process suggests a generic way to define
spectral ranking by dominant eigenvectors on general graphs, by
applying the eigenprojector of the (possibly normalized) adjacency matrix to a
given boundary condition.

\bibliography{biblio}

\end{document}